\documentclass[%
twocolumn,secnumarabic,amssymb, amsmath, tightenlines,
nobibnotes, aps, 
prl,
]{revtex4-1}

\usepackage{bm}		
\usepackage{amssymb,amsmath}
\usepackage{graphicx}

\begin{document}

\title{High-precision test of Landauer's principle in a feedback trap}

\author{Yonggun Jun}

\altaffiliation{Present address:  Department of Developmental and Cell Biology, University of California, Irvine, CA  92697-2300, USA}

\author{Mom\v cilo Gavrilov}
\thanks{The first two authors contributed equally to this work.}

\author{John Bechhoefer}
\email[email: ]{johnb@sfu.ca}

\affiliation{Department of Physics, Simon Fraser Universitway, Burnaby, B.C., V5A 1S6, Canada}

\begin{abstract}
We confirm Landauer's 1961 hypothesis that reducing the number of possible macroscopic states in a system by a factor of two requires work of at least kT ln 2.  Our experiment uses a colloidal particle in a time-dependent, virtual potential created by a feedback trap to implement Landauer's erasure operation.  In a control experiment, similar manipulations that do not reduce the number of system states can be done reversibly.  Erasing information thus requires work.  In individual cycles, the work to erase can be below the Landauer limit, consistent with the Jarzynski equality.
\end{abstract}

\maketitle

In 1961, Rolf Landauer proposed a fundamental link between information theory and physical systems \cite{landauer61}:  erasing information in a macroscopic or mesocopic system is an irreversible process that should require a minimum amount of work, $kT \ln 2$ per bit erased, where $T$ is the system temperature and $k$ is Boltzmann's constant.  This work is dissipated into a surrounding heat bath.  At the time, the motivation was to understand the minimum power a computer requires to function.    Surprisingly, logical operations---the computations themselves---can be coded using  logically reversible gates that in principle can be realized in a thermodynamically  reversible operation, with no dissipation \cite{bennett73}.  But eventually, a computer's memory must be reset to its original state, and such an operation is, according to Landauer, inherently dissipative.  As the only inherently dissipative operation, it determines the theoretical minimum power required to run a computer.

Landauer's principle acquired further significance some years later, when Charles Bennett (and, independently, Oliver Penrose) noted that it resolves the long-standing threat to the second law of thermodynamics posed by \textit{Maxwell's demon}  \cite{penrose70,bennett82,leff03,maruyama09}.  In modern language, a demon acquires information about a system, lowering its entropy and raising its free energy, and then uses this acquired free energy to do work.  Unless some aspect of the demon's operation is dissipative, it will use the energy of the surrounding heat bath to do work, violating the second law of thermodynamics.  Szilard \cite{szilard29}, Brillouin \cite{brillouin51}, and others \cite{leff03} proposed that the measurement step is inherently dissipative.  However, as Bennett showed,  measurements can, in principle, be done without work \cite{bennett82}.  If measurements and calculations do not require work, the only other possibility consistent with the second law is that the erasure step, required to return the computer to its original state, is dissipative.  Thus, Landauer's principle resolves the paradox created by Maxwell's demon.  Although these theoretical arguments have persuaded most physicists, there has been persistent  skepticism from a variety of authors, within and without the physics community \cite{maroney05,kish12,hemmo12,norton13}.  The continuing controversy makes  clear experimental tests of Landauer's principle particularly important.

Landauer's principle remained untested for over fifty years.  Tests have recently become possible because of two key recent developments:  The first advance is conceptual---a method for estimating the work done on a particle and the heat dissipated by that particle that is based solely on the trajectory $x(t)$ and a knowledge of the potential $U(x,t)$.  In particular, it does not rely on measuring the minute amounts of heat $(\approx 10^{-21}$ J) involved in the erasure of a single bit of information.  The method was first proposed by Sekimoto \cite{sekimoto97,sekimoto10} and  tested, for example, by studying a colloidal particle in an aqueous medium \cite{blickle06}.  Extensions have led to a new field, the stochastic thermodynamics of small systems  \cite{seifert12,klages13}.  By focusing solely on the trajectory and the potential, one can isolate and measure the quantities of direct interest, removing the contributions of work and dissipation from ancillary devices---computer, camera, illumination, etc.---that are irrelevant to calculating the work done by the potential on the particle and the heat dissipated into the surrounding bath. 

The second advance is technical---the development of ways to impose user-defined potentials on a small particle undergoing Brownian motion.  One way, for example, uses the localized potential forces created by optical tweezers formed from a highly focused laser beam.  Then, either by shaping the beam by diffractive optical elements or by rapidly moving the beam between two or more locations, a more complicated potential, such as a double well, can be created.  Such an approach was recently used to make a first test of Landauer's principle, under partial-erasure conditions \cite{berut12}.  A related approach \cite{toyabe10} had earlier been used to explore a Szilard engine \cite{leff03} that converts information to work, a process that may be regarded as an indirect test of Landauer's principle. 

\begin{figure}
	 \begin{center}
	 \includegraphics[width=4.5cm]{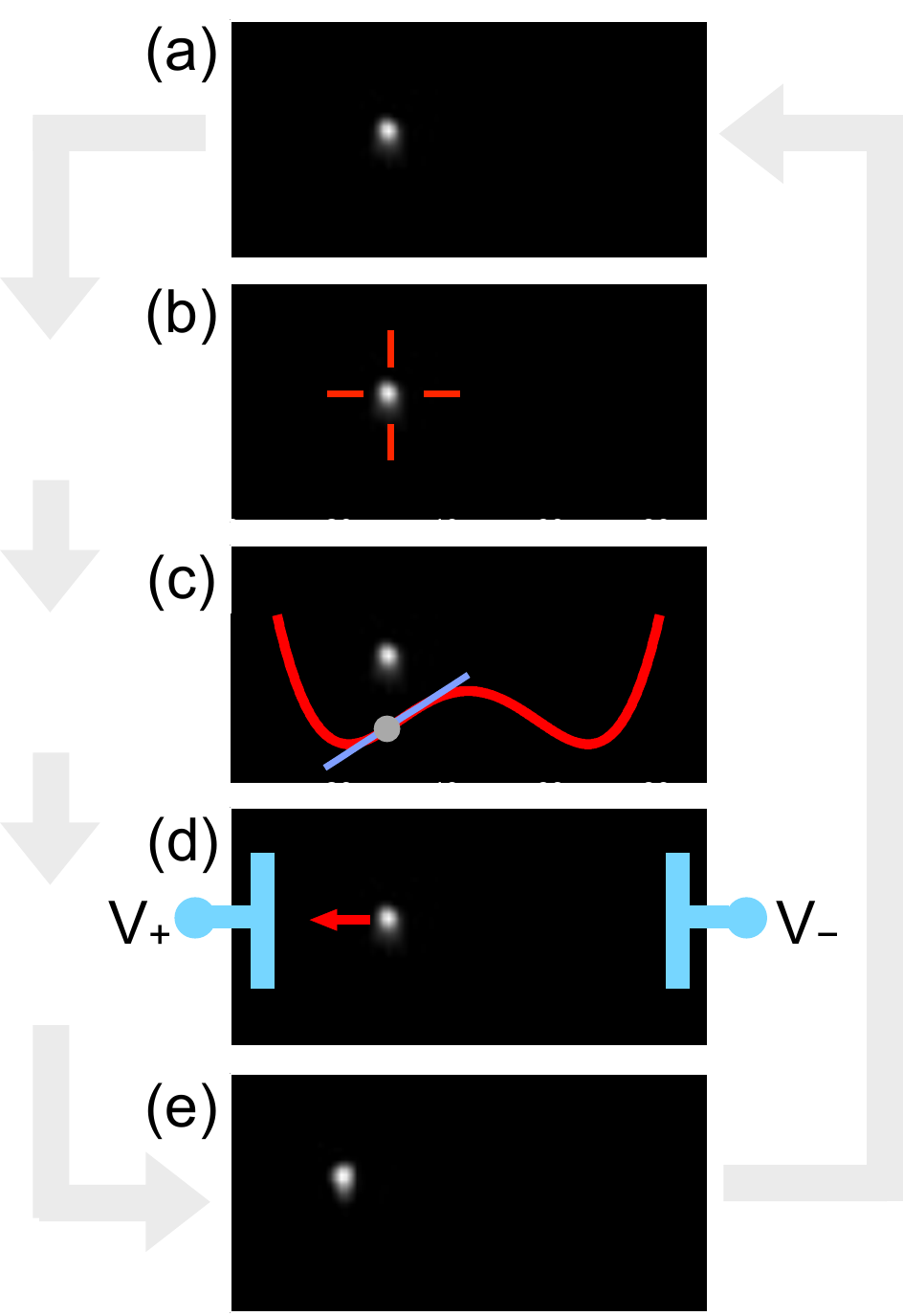}
	 \end{center}
	 \caption[example] { \label{fig:FBtrap} 
(color online). Schematic of feedback trap operation.  (a) Acquisition of an image of a fluorescent particle.  (b) Determination of particle position from that image using a centroid algorithm. (c) Evaluation of feedback force   $F_x=-\partial_x U(x,t)$ at the observed position $\bar{x}$. (d) Application of electric force, with voltage set by electrodes (light blue), held constant during the update time $t_s= 10$ ms. (e)  Acquistion of another image, with particle at a new position. The position of the particle is shifted relative to the previous one because of the deterministic feedback displacement due to the applied electric field, the stochastic effects due to  thermal diffusion, and observational noise due to the finite number of photons collected.  In the full trap, two sets of electrodes allow for two-dimensional trapping.  Using a thin sample cell (800 nm) confines the particle in the third dimension.}
\end{figure}

Here, we adopt a more flexible approach that uses feedback loops to create a virtual potential.  We implement a version of the Anti-Brownian Electrokinetic (ABEL) feedback trap to test Landauer's principle.  As illustrated in Fig.~\ref{fig:FBtrap}, the trap acquires an image of fluorescent particle diffusing in an aqueous solution and uses image processing to estimate its position.  Rather than using a physical potential, such as that produced by optical or magnetic tweezers, in a feedback trap, voltages are applied across two sets of electrodes, creating an electrical force that moves the particle.  In the present study, the force is chosen to create a \textit{virtual potential} $U(\bar{x}_n$) that is a discrete approximation to a corresponding continuous potential $U(x)$ \cite{cohen05b,cohen05d,jun12,gavrilov13}.  The potential is virtual because it is imposed by the feedback loop and is based on the measured position at $\bar{x}_n$, rather than the corresponding (unknown) actual position, $x_n$.  The feedback trap allows the  exploration of particle dynamics in a nearly arbitrary potential, where the only constraint is that the local relaxation time of the potential be significantly greater than the update time of the feedback loop.  The local relaxation time is the time for a particle to relax in a potential of curvature $\kappa(x)$, where $x$ is the local position of the particle \footnote{See Supplemental Material}.  A virtual potential may also be time dependent $ U(\bar{x}_n, t_n)$; the time variations must also be slow compared to the feedback update time $t_s$.  If the feedback update time is short, the dynamics in virtual potentials asymptotically approach those of the corresponding continuous potential \cite{jun12}.

In our experimental setup, described in detail in \cite{gavrilov13}, a camera takes images of 200-nm fluorescent particles in an inverted epifluorescence microscope.  Particles diffuse in two dimensions, as the 800-nm thick cell limits vertical motion.  For inserting the electrodes, the cell has two pairs of holes that are roughly orthogonal.  Fluorescent particles are illuminated by a 5-mW, 532-nm laser.  The computer estimates a particle's position in a camera image using a modified centroid algorithm and then generates a feedback force based on the inferred position.  The feedback force is applied as an electric force that is generated by applying voltage differences to two pairs of electrodes. The feedback loop is updated every $t_s = 10$ ms, with force updates at the middle of a 5-ms camera exposure.  The delay between observing a position and applying the calculated force is also 10 ms.

Since the feedback forces are generated by applying voltage differences to two pairs of electrodes, one must translate the voltages into forces properly.  The two sets of quantities are connected by a $2 \times 2$ \textit{mobility} matrix $\bm{\mu}$ that relates the two nominally orthogonal potentials to $x$ and $y$ displacements \cite{Note1}.  In the feedback trap, displacements are affected by slow drifts, most likely due to voltage offsets created by electrochemical reactions at the electrodes and by the voltage amplifier.  Removal of these drifts is essential for quantitative measurement of thermodynamic parameters such as work.  Here, we  estimate and correct for them in real time, using a recursive maximum likelihood (RML) method \cite{Astrom2008} for relating displacements to applied voltages.  The RML method gives an unbiased estimate of particle's properties, which are further used for imposing the ``virtual'' potential and measuring work with the high precision.  See \cite{gavrilov14} for a full discussion.

The ability to measure work with the high precision and the flexibility to choose the potential in a feedback trap gives it a key advantage in  testing Landauer's principle.  Previous tests, based on the rapid manipulation of optical tweezers, did not have full control of potential shape \cite{berut12}.  As a result, they were unable to achieve complete erasure, and corrections were necessary to connect to the $kT \ln 2$ result predicted by Landauer for the full erasure of one bit of information.  Follow-up studies used the Jarzynski relation to infer the Landauer value from finite-time cycles \cite{berut13} and explored the energetics of symmetry breaking \cite{roldan14}.  Here, a higher barrier prevents spontaneous hops across the barrier, ensuring complete erasure and approach to the limiting value of work, $kT \ln 2$.  Equally important, we are also able to perform a control experiment where, using similar manipulations in the potential that are chosen so as not to compress the phase space, the required work goes to zero at long cycle times, consistent with a reversible operation.  We thus directly show the link between phase-space compression and loss of reversibility.

\begin{figure}
	 \begin{center}
	 \includegraphics[width=5.5cm]{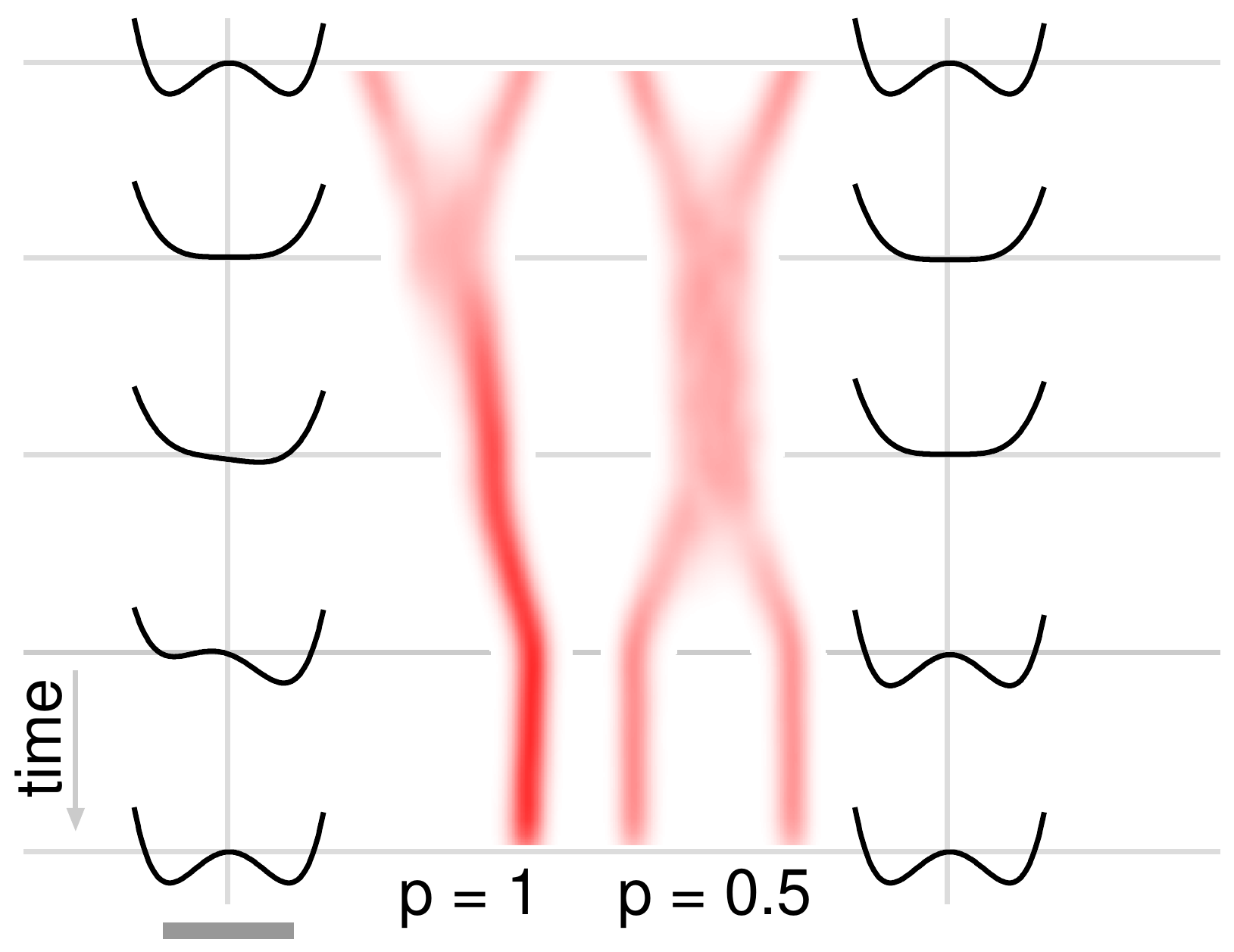}
	 \end{center}
	 \caption { \label{fig:trajectories} 
(color online).  Erasure protocol and trajectories for full erasure ($p=1$) and no erasure ($p=0.5$).  Full erasure requires a strong tilt of the potential towards the desired well ($A=0.5$).  In the no-erasure protocol, the potential is symmetric at every time step ($A=0$), implying that a particle ends up in a random final state.  The image intensity $I(x,t) \propto P(x,t)$ the occupation probability for a particle in a time-dependent, double-well potential and was generated from 30 trajectories for each case using kernel density estimation.  We used a Gaussian kernel with standard deviation equal to $0.1$ in time and $0.15~\mu$m in space, evaluated on a $500 \times 160$ grid.  Scale bar at lower left measures 5 $\mu$m.}
\end{figure}

Figure~\ref{fig:trajectories} illustrates the two protocols that we used.  At left is the \textit{full-erasure protocol}, denoted $p=1$ to indicate that the probability that a particle ends up in the right well, regardless of its initial state (left well or right) is unity.  Our erasure protocol is a modified version of that presented by Dillenschneider and Lutz \cite{dillenschneider09}.  The cyclic operation has four stages:  lower the barrier, tilt, raise the barrier, untilt.  To create the  protocol, we impose
\begin{equation}
	U(x,t) = 4 E_b \left[ -\tfrac{1}{2} g(t) \tilde{x}^2 + \tfrac{1}{4} \tilde{x}^4 - A f(t) \tilde{x} \right] \,,
\label{eq:DWpotential}
\end{equation}
where $\tilde{x} = x/x_m$ and where the energy barrier $E_b$ separates two wells located at $\pm x_m$.   The functions $g(t)$ and $f(t)$ control barrier height and tilt, respectively \cite{Note1}.  The full potential is harmonic in the transverse direction:  $U_{\rm tot}(x,y,t) = U(x,t)+\tfrac{1}{2}\kappa_y y^2$.

The associated density plot, where red intensity is proportional to the occupation probability $P(x,t)$ of the particle, shows that all trajectories end up in in the right well.  Figure~\ref{fig:trajectories} shows at right the \textit{no-erasure protocol}, which differs from the full-erasure protocol only in that there is no tilt.  From the symmetry, we expect (and experimentally we confirm) that there is an equal probability for particles to end up in the left or right well.  In this case, no net erasure of information occurs:  the particle has two possible states before and two possible states after the cyclic operation, and a further measurement would be needed to know the state of the system. 

The duration of each protocol (cycle period) is denoted $\tau$ and is measured in units of $\tau_0 = (2x_m)^2/D$, the time scale for particles to diffuse between wells at $\pm x_m$.  In the experiment, $x_m = 2.5$ $\mu$m, $D \approx 1.7$ $\mu$m$^2$/s, implying $\tau_0 \approx 15$ s.  The energy barrier $E_b$ is set to 13 $kT$.  Such a barrier height insures that the time between spontaneous hops (dwell time) is two orders of magnitude longer than the longest erasure cycle \cite{Note1}. The update time $t_s=10$ ms is fast enough that the discrete dynamics and work measurements are accurate estimates of the continuous equivalents for our set of parameters \footnote{M. Gavrilov, Y. Jun, and J. Bechhoefer, unpublished.}.

\begin{figure}[t]
	 \begin{center}
	 \includegraphics[width=8cm]{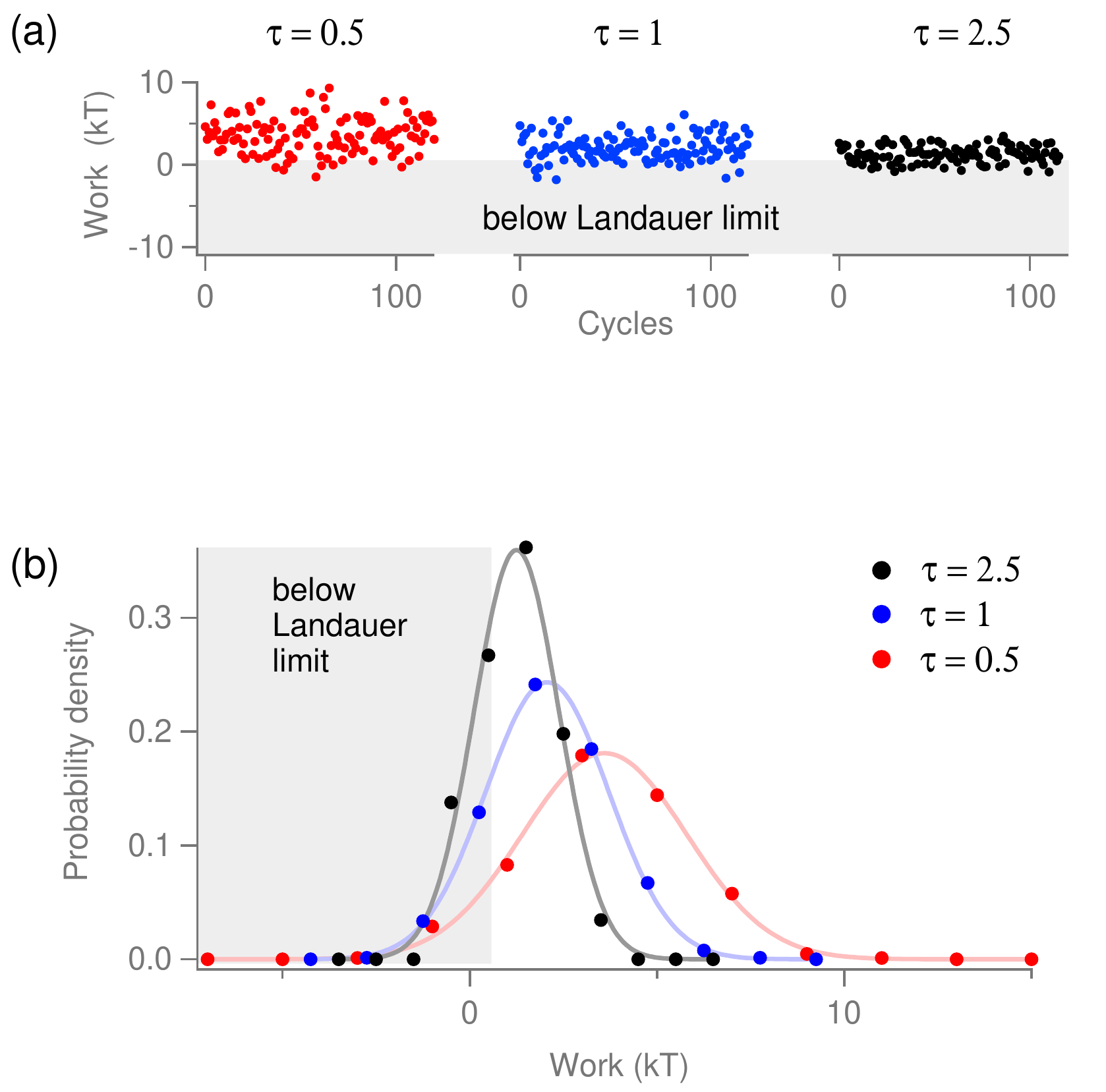}
	 \end{center}
	 \caption{
(color online). (a)  Experimental series of work values for individual cycles of duration $\tau =  0.5$, $1$, and $2.5$.  The gray shaded area shows those cycles where the measured work is below the Landauer limit.  (b)  Histograms of work series in (a), with Gaussian fits shown as solid curves.  The gray shaded area shows the part of the probability distribution that is below the Landauer limit.}
\label{fig:workSeries} 
\end{figure} 

To find the work in one erasure cycle, we evaluate the imposed potential $U[x(t),t]$ at the position of the particle and discretize Sekimoto's formula \cite{sekimoto97}, $W=\int_0^\tau dt \, (\partial_t U)$, where $\tau$ is the erasure cycle time.  We then have
\begin{equation}
	W (\tau) = - 4 E_b \sum_{n=0}^{N_s}  
	\left[ \tfrac{1}{2} (\Delta g)_n \, \tilde{x}_n^2  + A (\Delta f)_n \, \tilde{x}_n \right] \,,
\label{eq:WorkCalculation}
\end{equation}
where $\Delta g_n \equiv  \dot{g}(t_n) \, t_s$ and $\Delta f_n \equiv \dot{f}(t_n) \, t_s$ and $N_s$ is the number of steps in the erasure cycle.

Figure~\ref{fig:workSeries} shows that, for fixed $\tau$, the work in each cycle is stochastic, with $W(\tau)$ empirically distributed as a Gaussian random variable.  We estimate the mean work $\langle W \rangle_\tau$ for cycles of time $\tau$ by averaging over $N$ measurements.  From Fig.~\ref{fig:workSeries}, the standard error of the mean depends only on the total time $t_{\rm tot}$ taken by the $N$ cycles.  To keep the standard error of the mean constant for different cycle times, we thus choose $N(\tau) = t_{\rm tot} / \tau$.  With $t_{\rm tot}= 60$ min., the statistical errors were about $\pm 0.10$ $kT$.  To find the asymptotic work, we extrapolate results from finite-$\tau$ cycles to infinite cycle times via the expected $\tau^{-1}$ finite-time correction \cite{schmiedl08,sekimoto10,aurell12}:

\begin{equation}
	\frac{\langle W \rangle_\tau}{kT} = \frac{\langle W \rangle_\infty}{kT} + a  \tau^{-1} \,,
\label{eq:work-finite}
\end{equation}
where $\langle W \rangle_\infty / kT$ is $\ln 2$ for the full-erasure and 0 for the no-erasure protocols.

\begin{figure}
	 \begin{center}
	 \includegraphics[width=9cm]{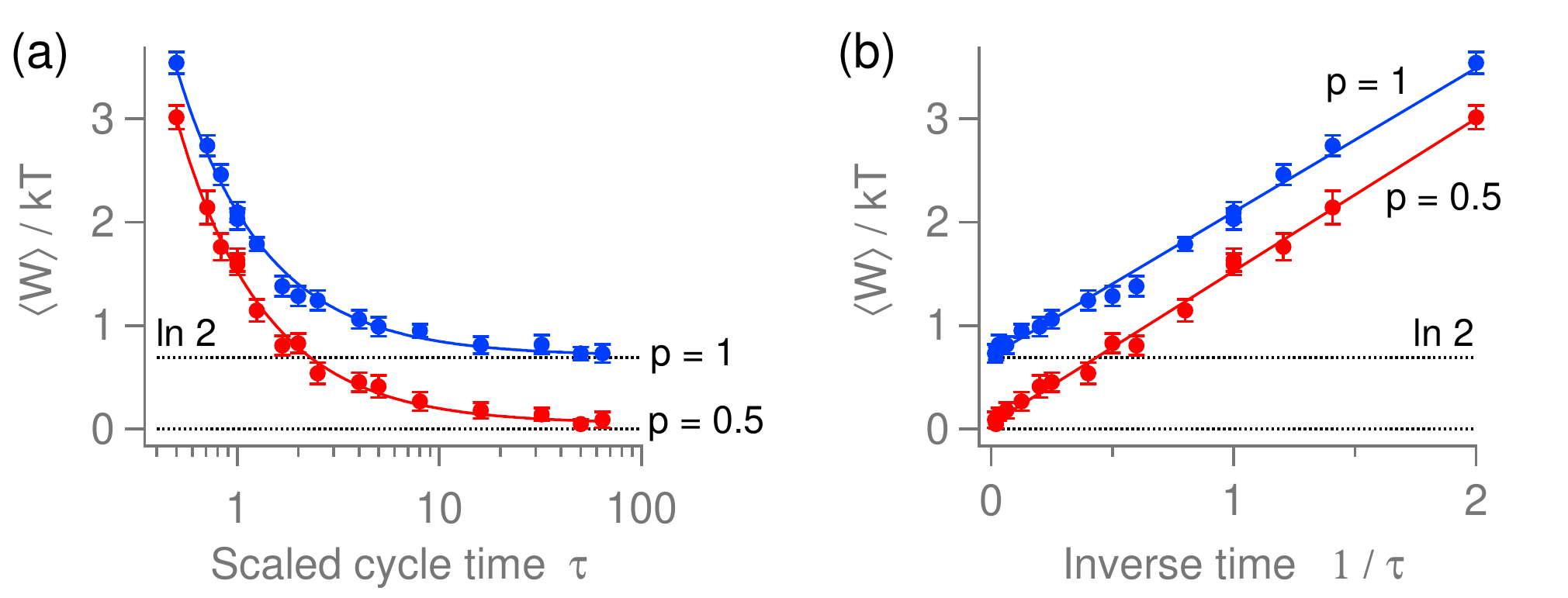}
	 \end{center}
	 \caption[example] { \label{fig:Work} 
(color online).  Mean work measured in the full erasure ($p=1$) and no-erasure ($p=0.5$) protocols.  (a) Mean work approaches the Landauer limits for each protocol. Solid line shows fit to  asymptotic $\tau^{-1}$ correction.  (b) Mean work as a function of inverse time.  The Landauer limit is given by the $y$-axis offset.  The dimensionless cycle times $\tau$ are in units of $\tau_0 = (2x_m)^2/D$.}
\end{figure} 

Figure~\ref{fig:Work} presents the main results of this study:  Part A shows the average work as a function of $\tau$ for both the full-erasure and no-erasure protocols.  The solid lines are fits to Eq.~\ref{eq:work-finite}.  To show the asymptotic form and its limit more clearly, we replot the data as a function of $\tau^{-1}$ in Fig.~\ref{fig:Work}B.  The  fit parameters are collected in Table \ref{table:work}.

\begin{table}
\begin{center}
\begin{tabular}{ l | c c c }
  & \begin{tabular}[c]{@{}c@{}}Asymptotic\\ work ($\pm 0.03$)\end{tabular} & \begin{tabular}[c]{@{}c@{}}Scale\\ time $a$\end{tabular} & \begin{tabular}[c]{@{}c@{}}$\chi^2$\\ $\nu = 14$\end{tabular} \\
\hline \\[-9pt]
full erasure ($p=1$) & 0.71 & 1.39 & 8.2 \\[2pt]
no erasure ($p=0.5$)  & 0.05 & 1.48 & 7.5 \\
\end{tabular}
\end{center}
\caption{Summary of results for full-erasure and no-erasure protocols.  Work $\langle W \rangle$ is divided by a factor of $kT$.  The full-erasure value is compatible with $\ln 2 \approx 0.693.$  The dimensionless parameter $a$, defined in Eq.~\ref{eq:work-finite},   gives the erasure scale time, in units of $\tau_0 = (2x_m)^2/D$.}
\label{table:work}
\end{table}

The asymptotic work values that we find are compatible with the expected values, $\ln 2 \approx 0.69$ and $0$, respectively; the dimensionless scale times $a$ are of order unity, also as expected; and the $\chi^2$ statistics indicate good fits.  Thus, we have shown experimentally that the full-erasure protocol, which involves the compression of phase space from two macroscopic states to one, asymptotically requires $kT \ln 2$ of work, while the very similar no-erasure protocol, which has no such phase-space compression, is reversible.  

The Gaussian work distributions seen in Fig.~\ref{fig:workSeries} have a mean that exceeds Landauer's limit in the erasure experiment.  However, individual cycles may have values of the stochastic work that are below the Landauer limit.  Indeed, they can even be  negative, drawing energy from the bath, in an apparent violation of the second law \cite{wang02}.   As a further check on our results, we note that when work distributions are Gaussian,  the Jarzynski equality---in this case, equivalent to linear response theory---implies a relation between the mean $\langle W \rangle$ and variance $\sigma_W^2$ of the work distribution [see, for example, Ref.~\cite{jarzynski11}, Eq.~21]:
\begin{equation}
	\sigma_W^2 = 2 \left( \langle W \rangle - \Delta F \right) \,,
\label{eq:jarzynski}
\end{equation}
where $\langle W \rangle$, $\Delta F$, and $\sigma_W$ are all measured in units of $kT$ and where $\Delta F$ is interpreted as a non-equilibrium free energy \footnote{
The non-equilibrium free energy change $\Delta F=\Delta E- T \, \Delta S$, with the stochastic system entropy $S = -k\sum_i p_i \ln p_i$.  We normalize $S$ by $k$, making it dimensionless.  The sum is over all states in the system (here, two states---particle in left well or in right well).  In our case, $\Delta E = 0$ because we consider cyclic operations, but $\Delta S = S_{\rm final} - S_{\rm initial} = 0$ for the no-erasure protocol (since, at the end of the cycle, $p_0 = p_1 = 0.5$ implies $S_{\rm final} = S_{\rm initial} = \ln 2$) and $\Delta S = -\ln 2$ for the full-erasure protocol (since, at cycle end, $p_0 = 0$ and $p_1 = 1$ implies $S_{\rm final} = 0$).}
, equal to ln 2 for the full-erasure protocol and 0 for the no-erasure protocol.

\begin{figure}[h!]
	 \begin{center}
	 \includegraphics[width=4.5cm]{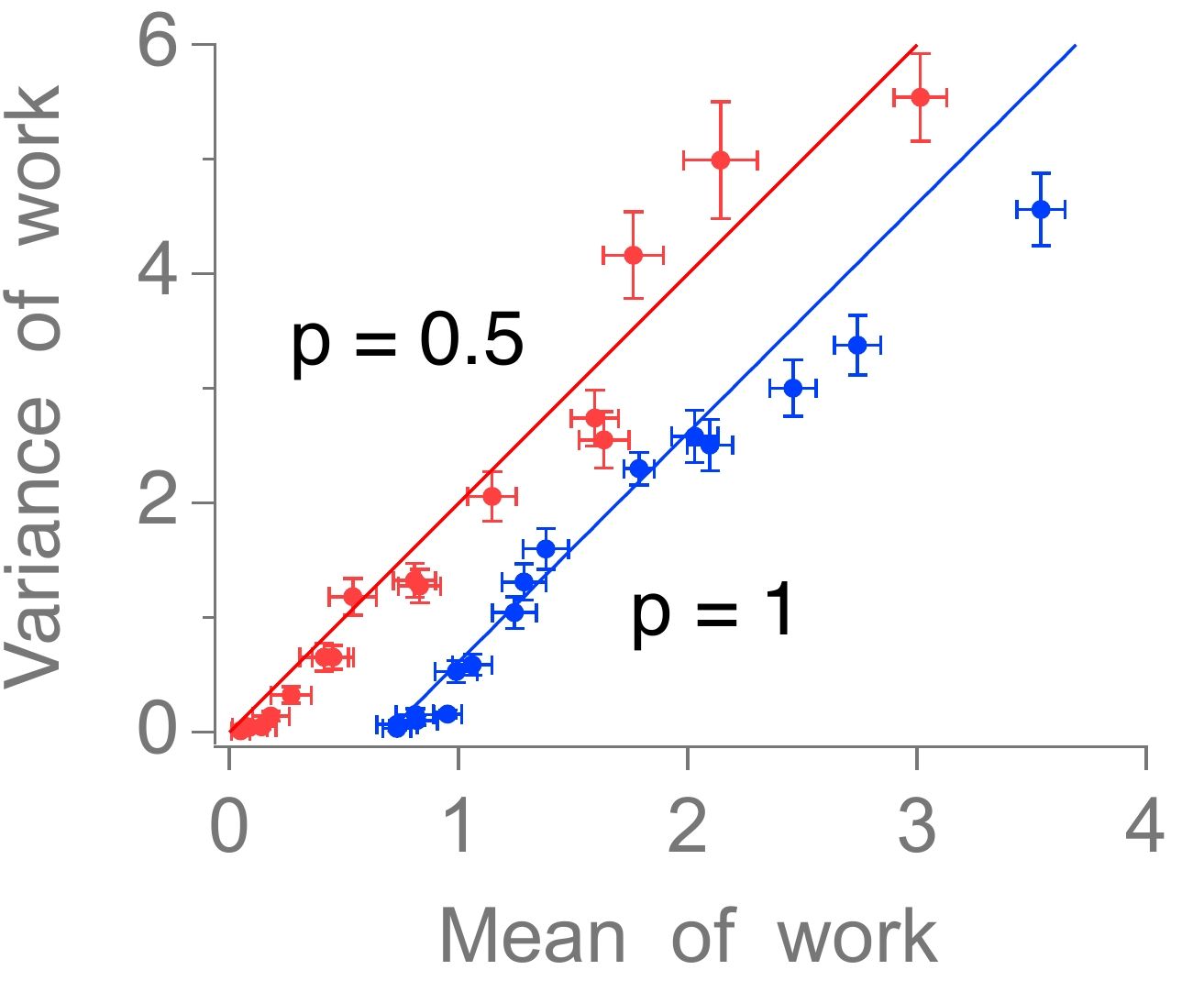}
	 \end{center}
	 \caption[example] { \label{fig:jarzynski} 
(color online). Variance versus mean for the work distribution, in units of $kT$.  Solid lines are plotted from the Jarzynski relation, Eq.~\ref{eq:jarzynski}, and have slope 2.}
\end{figure}

Figure~\ref{fig:jarzynski} shows these quantities for both protocols.  The solid lines are plots (not fits) from Eq.~\ref{eq:jarzynski}.  There is good agreement for longer cycle times that becomes poorer for shorter cycles, which have larger mean work values.  The shorter cycle times are problematic, both because the asymptotic result, Eq.~\ref{eq:jarzynski}, and the approximation of a virtual to a real potential can break down.  The Jarzynski equality has been explored in more detail in the context of Landauer's principle in Ref.~\cite{berut13}.  In combination with the expected decrease in mean work as cycle times are lengthened, it explains immediately why the work distribution sharpens for long $\tau$.  Because the variance of the mean estimate and the mean itself (see Eq.~\ref{eq:jarzynski}) both decrease as $\tau^{-1}$, measuring for a time $T$ leads to the same error-bar estimates, independent of the chosen cycle time $\tau$,  assuming that $\tau$ is long enough that the distributions are indeed  Gaussian.  For shorter cycle times, the distributions are expected to be non-Gaussian  \cite{dillenschneider09}.

In conclusion, the results presented here give the first direct test of Landauer's principle to confirm the predicted erasure energy of ln 2 per bit, connecting the seemingly disparate ideas of information and heat flow.  The tests also answer the threats to the second law of thermodynamics posed by Maxwell's demon.

The high precision and great flexibility of feedback traps opens doors to many further tests of stochastic thermodynamics and nonequilibrium statistical physics.  For example, Aurell et al. \cite{aurell12} have studied the problem of  minimizing the work in finite-time operations and concluded that optimal protocols involve complex, discontinuous manipulations of potentials that would be hard to impose in any other way.  The potential from a feedback trap need not even come from a potential, making possible the exploration of non-potential dynamics \cite{cohen05d}.  Feedback traps are, of course, also natural settings for exploring non-equilibrium extensions of the Landauer theory \cite{esposito11,sagawa14}, as well as connections between feedback and thermodynamics \cite{seifert12,klages13}.

Landauer's link between information theory and physical systems is critical for understanding performance limits in nanoscale machines and biological systems.  At present, the lowest energies involved in elementary operations in computation such as switching are $\approx 1000~kT$ \cite{pop10}, which are approaching the energy scales ($10$--$100~kT$) used by  biological systems to sense the outside world and make decisions \cite{lan12}.  At these energy scales, the fundamental explorations of equilibrium and non-equilibrium systems made possible by the new methods used here will become increasingly important.

We thank Massimiliano Esposito, Suckjoon Jun, and David Sivak for helpful discussions. The sample cell was fabricated in the 4D Labs facility at Simon Fraser University.  This work was supported by NSERC (Canada).

\bibliography{sample}

\newpage
\section*{Supplemental Material:}
\section{Erasure protocol}
\label{sec:erasure}
Our erasure protocol is similar to that presented by Dillenschneider and Lutz \cite{dillenschneider09} but separates explicitly the operations that change the barrier height and tilt the potential.  The function $g(t)$ in Eq.~1 lowers and raises the barrier.  A lowered barrier ($g \approx 0$) allows a particle to explore both states on a time scale set by diffusion, $\tau_0 \equiv (2x_m)^2/D$.  The time scale $\tau_0$ sets the basic scale for trap dynamics:  achieving full erasure requires a cycle time $\tau \gg \tau_0$.  For this reason, we scale times by $\tau_0$ (rather than by $t_s$, the time per update step).
\begin{figure}[h!]
	 \begin{center}
	 \includegraphics[width=6cm]{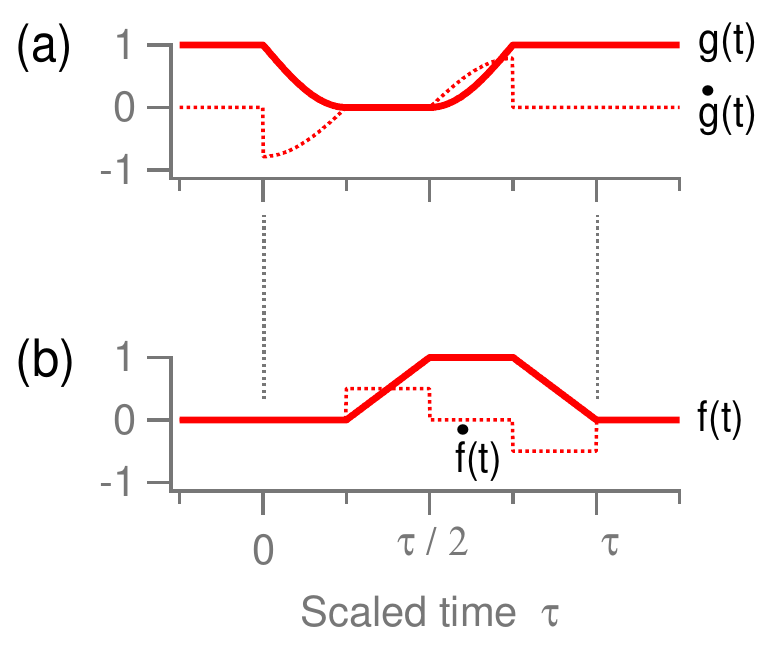}
	 \end{center}
	 \caption{Control functions. (a) Barrier control $g(t)$ (solid red line) and its time derivative $\dot{g}(t)$ (dotted red line), used for the work calculation.  (b) Potential tilt $f(t)$ (solid red line) and its derivative $\dot{f}(t)$ (dotted red line).}
\label{fig:controlFuncs} 
\end{figure}

The linear term $Af(t)\tilde{x}$ in Eq.~1 tilts the potential, with $A$ setting the tilt amplitude and favoring one state and $f(t)$ the time-dependent protocol.  The functions $f(t)$ and $g(t)$ take values $\in [0,1]$ and are plotted in Fig.~\ref{fig:controlFuncs}.

\renewcommand{\arraystretch}{1.5}	%global vertical padding to make table look better
\begin{table}[th!]
\begin{center}
\begin{tabular}{l|l|l|l}
	time	& $g(t)$	& $f(t)$	& action \\ \hline
	$t \in[0, \tfrac{1}{4} \tau]$  & $1-\sin[\omega(t)]$   & 0  & lower barrier  
    		\\ \hline
	$t \in[\tfrac{1}{4}\tau, \tfrac{1}{2}\tau]$ & $0$ & $4t/\tau -1$ & tilt  
    		\\ \hline
	$t \in[\tfrac{1}{2}\tau, \tfrac{3}{4}\tau]$ & $1-\sin[\omega(t-\tau/4)]$           
    			& 1 &  raise barrier  
    		\\ \hline
	$t \in[\tfrac{3}{4}\tau, \tau]$  & $0$  & $4 - 4t/\tau$ & untilt    
    		\\ \hline
	$t \notin[0, \tau]		       $  & $1$           & $0$ & static double-well   
\end{tabular}
\end{center}
\caption{Protocols for lowering and raising the barrier ($g$) and for tilting the potential ($f$).  Time $t$ and cycle time $\tau$ are both scaled by the diffusion time $\tau_0$, and $\omega = 2\pi/\tau$.}
\label{table:protocols}
\end{table} 

Table~\ref{table:protocols} defines explicitly the four stages of the protocol, each  lasting a quarter cycle.  The stages are as follows:  First, lower the barrier; then tilt the potential to favor one state; then lock the particle in that state by raising the barrier; and, finally,  return the potential to its initial form by untilting it.  

\section{Barrier heights and dwell time}
\label{sec:barrier}

For the symmetric double-well potential, it is important that the barrier be so high that  the probability for a spontaneous ``hop" across is vanishingly small on the time scale of the longest cycle time explored.  Such hops would imply an incomplete erasure and thus a work that is less than the  $(kT$ ln $2)$ limit.  In previous experiments\cite{berut12, berut13}, the barrier was low enough that spontaneous hops occurred.  Erasure was thus incomplete and corrections for the finite barrier height had to be made.  
\begin{figure}[h!]
	 \begin{center}
	 \includegraphics[width=9cm]{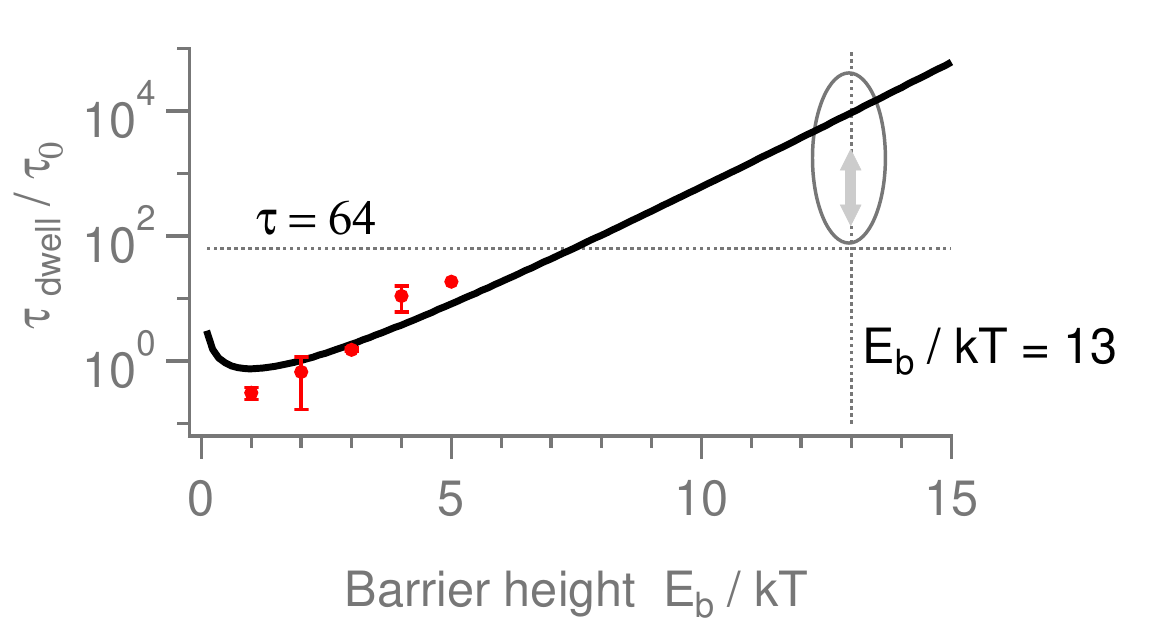}
	 \end{center}
	 \caption[example] { \label{fig:KramersSimulation} 
Mean Kramers times for different barrier heights. Solid red markers indicate experimental data points.  Black curve shows the Kramers theory, Eq.~\ref{eq:KramersSpecial}, for double-well potential with $x_m = 2.5$ $\mu$m and $E_b=$ 13 $kT$.  The vertical dotted line shows the barrier height for the erasure experiment.  The horizontal dotted line shows the longest erasure cycle time.  The circled short, vertical line with double arrows represents the factor of $\approx 100$ between the predicted dwell time and the longest experimental cycle time.}
\end{figure}

We thus explored the relation between barrier height, $E_b/kT$ and \textit{dwell time}, the mean time between spontaneous ``hops" across the barrier.  Figure~\ref{fig:KramersSimulation} gives dwell time vs.~barrier height, showing theory (solid line), and experiment (red solid markers).  Although we are interested in long dwell times in the erasure experiment, we can directly measure the dwell time only for lower barriers, where the dwell times are accessible experimentally,  as in previous work \cite{cohen05b}.  The theoretical predictions for dwell time are based on Eq.~\ref{eq:KramersSpecial}.  The relevant  experimental point is that at the chosen barrier height, $E_b/kT = 13$, the expected dwell time is about two orders of magnitude greater than the longest experimental cycle time (circled vertical line with double arrows).  This separation of time scales implies full erasure in a double-well potential.

The theoretical prediction for the dwell time is derived from the overdamped limit of Kramers theory \cite{hanggi90},

\begin{align}
	\left(\frac{\tau_{\rm dwell}}{\tau_0}\right) 
	&= \frac{2\pi}{\tau_0}\left(\frac{1}{\sqrt{|\kappa_0\kappa_m|}}\right)
				\left(\frac{1}{D}\right) e^{E_b} \nonumber \\[3pt]
	&=  \left( \frac{\sqrt{2}\pi}{16}\right) \left( \frac{e^{E_b}}{E_b} \right) \,.
\label{eq:KramersSpecial}
\end{align}

In the second expression, we substitute $\tau_0 = (2x_m)^2/D$ and insert the curvatures of the potential at the barrier and the well:  $|\kappa_0| = |\partial _{xx}U(x=0)| = 4E_b/x_m^2$ and $|\kappa_m| = |\partial _{xx}U(x=x_m)| = 8E_b/x_m^2$, respectively.   Note that the dwell time diverges for $E_b/kT \to 0$.  This surprising conclusion, just noticeable as the upturn in the solid curve in Fig.~\ref{fig:KramersSimulation}, is an artifact of the calculation itself, which assumes $E_b/kT \gg 1$.

\end{document}